\documentclass[preprint,nofootinbib]{revtex4}

\usepackage{amsmath}
\usepackage{graphicx}

\newcommand{\axif}{\left|\xi_f\right|}

\newcommand{\xib}{\overline{\xi}}
\newcommand{\xibh}{\widehat{\xib}}
\newcommand{\xibf}{\overline{\xi}_f}

\newcommand{\axibf}{\left|\xibf\right|}
\newcommand{\axibh}{\left|\widehat{\xib}\right|}
\newcommand{\axibhf}{\left|\widehat{\xib}_f\right|}
\newcommand{\su}{SU(3)}
\newcommand{\ket}[1]{\left|#1\right\rangle}

\newcommand{\braket}[2]{\left\langle #1 | #2 \right\rangle}

\newcommand{\K}{K^0}
\newcommand{\Kb}{\overline{K}{}^0}

\newcommand{\lamsc}{V^*_{cb}V_{cs}}
\newcommand{\lamsu}{V^*_{ub}V_{us}}
\newcommand{\lamdc}{V^*_{cb}V_{cd}}
\newcommand{\lamdu}{V^*_{ub}V_{ud}}

\renewcommand{\Re}{\mathcal{R}e}
\renewcommand{\Im}{\mathcal{I}m}
\newcommand{\abs}[1]{\left\vert#1\right\vert}
\newcommand{\norm}[1]{\left\Vert#1\right\Vert}

\begin{document}

\preprint{{\vbox{\hbox{}\hbox{}\hbox{}
\hbox{hep-ph/0508046}}}}

\vspace*{1.5cm}

\title{Using SU(3) Relations to bound the CP Asymmetries in
  $B\to KKK$ decays}

\author{Guy Engelhard}\email{guy.engelhard@weizmann.ac.il}
\affiliation{Department of Particle Physics,
  Weizmann Institute of Science, Rehovot 76100, Israel}
\author{Guy Raz}\email{guy.raz@weizmann.ac.il}
\affiliation{Department of Particle Physics,
  Weizmann Institute of Science, Rehovot 76100, Israel}


\vspace{2cm}
\begin{abstract}
  We consider three body $\Delta s = 1$ $B\to f$ decays with $f=KKK$. The
  deviations of $-\eta_f S_f$ from $S_{\psi K_S}$ and of $C_f$ from zero can
  be bounded using the approximate $\su$ flavor symmetry of the strong
  interactions and branching ratios of various $\Delta s = 0$ modes. We
  present the most promising $\su$ amplitude relations that can be used to
  obtain these bounds.
\end{abstract}

\maketitle

\section{Introduction}
\label{sec:introduction} Recently there has been a growing interest,
driven by experiments \cite{Abe:2005lp,Aubert:2005ha}, in the CP
asymmetries of $b\to s$ penguin dominated processes. For
final CP eigenstates the CP asymmetry is time dependent:
\begin{equation}
  \label{eq:1}
  \mathcal{A}_f(t) = \frac{\Gamma(\overline{B}^0_{}(t)\to
    f) - \Gamma(B^0_{}(t)\to f)}{\Gamma(\overline{B}^0_{}(t)\to
    f) + \Gamma(B^0_{}(t)\to f)} = -C_f\cos(\Delta
  m_B\,
  t) + S_f\sin(\Delta m_B\, t).
\end{equation}
For flavor specific states, the CP asymmetry is time independent:
\begin{equation}
  \label{eq:2}
  \mathcal{A}_f(t) = \mathcal{A}_f.
\end{equation}
Within the SM, these processes are dominated by a single phase,
while a second phase is CKM suppressed by $O(\lambda^2)$.
Consequently, within the SM, $-\eta_f S_f \approx S_{\psi K_S}$ and
$C_f,\mathcal{A}_f \approx 0$. Given the present central values and
errors of the experimental measurements
\cite{Abe:2005lp,Aubert:2005ha}, it is quite possible that new
physics is inducing shifts of $\mathcal{O}(0.1 - 0.4)$ from the SM
value. To understand whether new physics is indeed involved, it is
necessary to estimate the deviations allowed within the SM of the
various $-\eta_f S_f$'s from $S_{\psi K_S}$. Various methods have
been devised in order to estimate the deviations
\cite{Ali:1998eb,Beneke:2001ev,Beneke:2003zv,Bauer:2004tj,Buchalla:2005us,Beneke:2005pu}.
These method suffer from hadronic uncertainties and, furthermore,
are generally not suitable for use in three-body decays.

Recently, methods employing the $\su$ symmetry to place hadronic-model
independent bounds on CP asymmetries in $b\to s$ processes
\cite{Grossman:2003qp,Grossman:1997gr,Gronau:2003ep,Chiang:2003pm,Gronau:2003kx,Gronau:2004hp}
have been extended to three-body decay modes \cite{Engelhard:2005hu}.
The final state studied in \cite{Engelhard:2005hu}, $K_SK_SK_S$, is
symmetric under the exchange of any two of the final particles. This
situation allowed a considerable simplification of the $\su$ analysis.
In this work we extend the analysis to all other $KKK$ modes. Here
this exchange symmetry does not apply. A related study using
factorization to estimate the asymmetry in some of these modes is
presented in \cite{Cheng:2005ug}.

While the bounds we obtain are indeed hadronic-model independent, they
do suffer from two limitations. First, the $\su$ breaking is estimated
to be of $\mathcal{O}(0.3)$. The bounds we obtain may be violated at
this order. Second, since the $\su$ analysis cannot infer the phases
of physical amplitudes, we conservatively add the amplitudes coherently.
Generically, this means that our upper bounds can be considerably
weaker than the actual values.

The experimental data concerning CP asymmetries in $B\to KKK$ decays
is summarized in table~\ref{tab:BtoKKK}.
\begin{table}
  \centering
  \begin{tabular}{|c|c|c|c|}
    \hline
    Mode & $-\eta_f S_f$ & $C_f,\, -{\cal A}_f$ & Ref.\\
    \hline
    \hline
    $K_SK_SK_S$ & $0.26 \pm 0.34$ & $-0.41 \pm 0.21$ & \cite{HFAG} \\
    \hline
    $K^+K^-K_{S,L}$ & $0.51\pm 0.13$ & $0.08 \pm 0.09$ & \cite{Abe:2005lp} \\
    \hline
    $K_SK_SK_L$ & --- & --- & \\
    \hline
    \hline
    $K_SK_SK^\pm$ & n/a & $0.04 \pm 0.11$ & \cite{HFAG} \\
    \hline
    $K^\pm K^+K^-$ & n/a & $-0.02 \pm 0.08$ & \cite{HFAG} \\
    \hline
    $K_SK_LK^\pm$ & n/a & --- & \\
    \hline
  \end{tabular}
  \caption{Measured CP asymmetries in $B \to KKK$ decays.}
  \label{tab:BtoKKK}
\end{table}
The result of $S_{K^+K^-K_{S,L}}$ takes into account both
$S_{K^+K^-K_S}$ and $S_{K^+K^-K_L}$. The CP asymmetries of $B\to
K_SK_SK_L$ and $B^{\pm}\to K_SK_LK^\pm$ are not yet measured. For
completeness we include here also the result for $B\to K_SK_SK_S$
which was studied in~\cite{Engelhard:2005hu}.

The plan of the paper is as follows. In section
\ref{sec:constr-cp-asymm} we review the notations and formalism
relevant to three body decays. In section~\ref{sec:su-relations-bto}
we list the amplitude relations for $B\to KKK$ decays. We conclude in
section \ref{sec:conclusions}. We quote the experimental branching
ratios we use in appendix \ref{sec:exp-data}.
\section{Notations and Formalism}
\label{sec:constr-cp-asymm} In this section we review the
notations and formalism relevant to 3-body decays introduced in
\cite{Engelhard:2005hu} (where a much more detailed discussion is
presented). We use abstract vector notation, {\it e.g.}
$\vec{A}_f$, where the vector index runs over all possible values
for the quantum numbers, to describe the various states. The total
decay rate is given by
\begin{equation}
  \label{eq:3}
  \Gamma(B^0 \to f)=\left\|\vec{A}_{f}\right\|^2\;.
\end{equation}
Experiments measure the averaged rates given by:
\begin{equation}
  \label{eq:4}
\Gamma(B\to f)=\frac12\left[\Gamma(B^0\to
  f)+\Gamma(\overline{B}^0\to\overline{f})\right],
\end{equation}
where $\overline{f}$ is the CP-conjugate state of $f$.

The norm on the right hand side of~\eqref{eq:3} represents a sum over
all possible final states, that is, all momentum configurations. In
order to derive $\su$ relations, we choose to span the final states in
a basis with definite linear momenta. Our convention is that the order
in which we write the three final mesons corresponds to their momentum
configuration:
\begin{equation}
  \label{eq:5}
  \left|M_iM_jM_k\right\rangle\equiv
  \ket{M_i(p_1)M_j(p_2)M_k(p_3)}.
\end{equation}

We write a generic $\Delta s = 1$, $B^0\to f$ decay amplitude as follows:
\begin{equation}
  \label{eq:6}
  \vec{A}_{f}= \lamsc\, \vec{a}^c_{f} +
    \lamsu\, \vec{a}^u_{f}.
\end{equation}
Here, and for all other processes discussed below, the amplitudes for
the CP-conjugate processes, $\overline{B}{}^0\to \overline{f}$, have
the CKM factors complex-conjugated, while the $\vec{a}^{u,c}_f$
factors remain the same.

We define two parameters:
\begin{eqnarray}
  \label{eq:7}
  &\xi_f\equiv& \frac{\left|\lamsu\right|}{\left|\lamsc\right|}\,
  \frac{\vec{a}^c_{f} \cdot
    \vec{a}^u_{f}}{\|\vec{a}^c_{f}\|^2}\;,\\
  &\axibf \equiv& \frac{\left|\lamsu\right|}{\left|\lamsc\right|}\,
  \frac{\|\vec{a}^u_{f}\|}{\|\vec{a}^c_{f}\|}\;,
\end{eqnarray}
where
\begin{equation}
  \label{eq:8}
  \frac{\axif}{\axibf} \leq 1 \;.
\end{equation}
The parameter $\axibf$ is the one which can be constrained by $\su$
relations, and that leads, through eq. (\ref{eq:8}), to a constraint
on $\axif$. The CP asymmetries can be written to first order in
$\Re(\xi_{f})$ and $\Im(\xi_{f})$ as follows:
\begin{align}
  \label{eq:9}
  -\eta_{f}S_{f} - S_{\psi K_S} & = 2\, \cos 2\beta\, \sin
  \gamma\, \Re(\xi_{f})\;, \\
  \label{eq:10}
  -\mathcal{A}_f,\,  C_{f} & = - 2\, \sin\gamma\,
  \Im(\xi_{f})\;.
\end{align}

We write a generic $\Delta s=0$ decay amplitude as follows:
\begin{equation}
  \label{eq:11}
  \vec{A}_{f}=V^*_{cb}V_{cd}\, \vec{b}^c_{f}+V^*_{ub}V_{ud}\, \vec{b}^u_{f}\;.
\end{equation}
$\su$ symmetry leads to amplitude relations of the form
\begin{equation}
  \label{eq:12}
  \vec{a}^q_{f} = \sum\limits_{f'} X'_{f'}\,
  \vec{b}^q_{f'}\; \ \ \ (q=u\ {\rm or}\ c).
\end{equation}
Taking the norm of eq.~\eqref{eq:12} needs to be done with care: the
sum can involve states with different symmetry properties under
exchange of final particles since the $\Delta s = 1$ mode and the
$\Delta s = 0$ modes may have different symmetries with respect to the
momentum variables (so that integration over phase space is different
for the various modes). Taking the norm of both sides, the left hand
side can be bounded from above as:
\begin{equation}
  \label{eq:13}
  \left\|\vec{a}^q_{f}\right\| \leq \sum\limits_{f'}
  \left|X_{f'}\right|\,\left\|\vec{b}^q_{f'}\right\|\;,
\end{equation}
where $X_{f'}$ and $X'_{f'}$ are related by symmetry factors.

In order to bound $\axibf$ with no additional assumptions
\cite{Grossman:2003qp,Engelhard:2005hu}, we define another parameter,
$\axibhf$,
\begin{equation}
  \label{eq:14}
  \axibhf^2 \equiv \left|\frac{V_{us}}{V_{ud}}\right|^2
  {\tfrac{\left\| \lamdc\,  \vec{a}^c_{f}
        +\lamdu\,\vec{a}^u_{f} \right\|^2+\left\| V_{cb}V^*_{cd}\,
        \vec{a}^c_{f}
        +V_{ub}V^*_{ud}\,\vec{a}^u_{f} \right\|^2}
    {\left\| \lamsc\,  \vec{a}^c_{f}
        +\lamsu\,\vec{a}^u_{f} \right\|^2+\left\| V_{cb}V^*_{cs}\,
        \vec{a}^c_{f}
        +V_{ub}V^*_{us}\,\vec{a}^u_{f} \right\|^2}}\;.
\end{equation}
The numerator and denominator of $\axibhf^2$ are related to
charge-averaged rates:
\begin{align}
  \label{eq:15}
  \begin{split}
    \left\| \lamdc\, \vec{a}^c_{f}+\lamdu\,\vec{a}^u_{f} \right\|^2 &
    +\left\| V_{cb}V^*_{cd}\, \vec{a}^c_{f} +
      V_{ub}V^*_{ud}\,\vec{a}^u_{f} \right\|^2 \leq
    2\left(\sum\limits_{f'}\left|X_{f'}\right| \sqrt{\Gamma({B}\to
        f')}\right)^2\;,
  \end{split} \\
  \label{eq:16}
  \begin{split}
    \left\| \lamsc\, \vec{a}^c_{f}+\lamsu\,\vec{a}^u_{f}
    \right\|^2 & +\left\| V_{cb}V^*_{cs}\, \vec{a}^c_{f}
      +V_{ub}V^*_{us}\,\vec{a}^u_{f} \right\|^2 = 2\Gamma({B}\to f)\;.
  \end{split}
\end{align}
Using the measured charge-averaged rates, a constraint on
$\axibhf^2$ is obtained.

The $\axibhf$ and $\axibf$ parameters are related as follows:
\begin{equation}
  \label{eq:17}
  \axibhf^2 = \frac{ \left|\frac{V_{us}V_{cd}}{V_{cs}V_{ud}} \right|^2
    +\axibf^2 + 2\cos\gamma\,\Re\left(\frac{V_{us}V_{cd}}{V_{cs}V_{ud}}\,\xi_f\right)}{1 +
    \axibf^2 + 2 \cos\gamma\,\Re(\xi_f)}\;.
\end{equation}
The relation \eqref{eq:17} has the property that for $\lambda^2
\lesssim \axibhf\leq 1$ we get a constraint on $\axibf$, for any
$\xi_f$ (of course, within the allowed range, $|\xi_f|\leq\axibf$, see
eq.~\eqref{eq:8}). Since we do not know the value of $\xi_f$, we
should consider the weakest constraint, which corresponds to
$\Re(\xi_f)=\axibf$ (the $(V_{us}V_{cd})/(V_{cs}V_{ud})$ term is
experimentally known to be real to a good approximation). The weakest
bound, which corresponds to $\Re(\xi_f)=\axibf$ and $\gamma=0$, is
obtained from the curve $\axibhf=(\axibf-\lambda^2)/(1+\axibf)$.

In the following sections, we present the $\su$ analysis for the
$B\to KKK$ modes. Before presenting the relations, however, we make
two comments.

The first comment is related to the removal of resonances. In general,
the cleanest result would be obtained with all resonances removed both
from the CP asymmetries and from the branching ratios. Since, however,
we use $\su$ symmetry here, a resonance from a complete $\su$
representation, which enters all relevant modes, would not harm the
analysis as long as the same strategy regarding it is employed in all
branching ratios and in the CP asymmetry. One should also note that if
resonances are small in the $B\to KKK$ decays we consider (see for
example~\cite{Aubert:2005kd}), a removal of the resonances from the
$\Delta s=0$ branching ratio would only create a small error. On the
other hand, if the resonances are removed from the branching ratio of
the $B\to KKK$ mode, a failure to remove the resonances in the $\Delta
s=0$ modes would only weaken the bound but not invalidate it.

The second comment addresses our methodology of presenting the
results. In order to find the $\su$ relations of various $\Delta s =
1$ modes we follow the method outlined in \cite{Engelhard:2005hu}. We
scan over all possible contractions of the relevant $\su$ tensors,
avoiding the need to discuss $\su$ properties of tensor products. In
contrast to \cite{Engelhard:2005hu}, however, here we cannot use
symmetrized states only, a fact that complicates the analysis in the
following ways:
\begin{enumerate}
\item We have $92$ $\Delta s=0$ modes (including all possible
  permutations of the mesons) and $40$ reduced matrix elements. This
produces a table which is too large to include here.
\item Since there are $92$ modes, presenting the most general
  relations with free parameters (see e.g.
\cite{Grossman:2003qp}) is not practical.
\end{enumerate}
Since the contributions are added coherently, our constraints are
weakened if too many modes are included. Consequently the fewer the
modes involved in an $\su$ relation, the better the chance of
getting a strong bound. In the following section, we therefore
present only the relations involving the smallest number of modes,
hoping that (once measured) these will provide us with a strong
bound on the deviations.


\section{$\su$ relations for $B\to KKK$ decays}
\label{sec:su-relations-bto}

In this section we discuss the most promising $\su$ relations
relevant to each $B\to KKK$ mode. One can divide the $KKK$ states
into six distinct types which are determined by the identity of the
$K$'s and the symmetry of the states. States which are of the same
type have the same $\xi$ and related CP asymmetries. They are
related to each other by $K$ mixing factors. There are three types
which are common states of $B^0$ and $\overline{B}^0$: $K_SK_SK_S$
($K_LK_LK_L$), $K^+K^-K_S$ ($K^+K^-K_L$) and $K_SK_SK_L$
($K_LK_LK_S$). The other three types are charged and therefore
flavour specific: $K_SK_SK^\pm$ ($K_LK_LK^\pm$), $K^\pm K^+K^-$ and
$K_SK_LK^\pm$.

The first type was considered in~\cite{Engelhard:2005hu}, and the
others are considered below.

\subsection{$B\to K^+K^-K_{S,L}$}
\label{sec:su-relations-k+k}

While the mode $K^+K^-K_{S(L)}$ is, strictly speaking, not a CP
eigenstate, isospin~\cite{Abe:2002ms} and angular momentum
analysis~\cite{Aubert:2004ec} show that it is predominantly CP
even (odd).

Note that the $\Delta s = 1$ state $K^+K^-\K $ is related to the state
$K^+K^-K_{S,L}$ through $K$ mixing. We evaluate $\abs{\xi_{K^+K^-\K }}$,
but from eq. \eqref{eq:7} it is clear that $\abs{\xi_{K^+K^-\K }} =
\abs{\xi_{K^+K^-K_{S,L}}}$, where the latter is the quantity that enters
the bounds in eqs.~\eqref{eq:9} and~\eqref{eq:10}.

\subsubsection{Three $\Delta s=0$ Amplitudes}
We find two amplitude relations involving three $\Delta s=0$
modes:
\begin{eqnarray}
  \label{eq:18}
  \vec{a}^q_{K^+K^-\K } = \sqrt{\frac{3}{2}}\,\vec{b}^q_{K^+K^-\eta_8} -
  \frac{1}{\sqrt{2}}\,\vec{b}^q_{K^+K^-\pi^0} +
  \vec{b}^q_{\pi^+K^-\K }\;,\\
  \vec{a}^q_{K^+K^-\K } = \sqrt{\frac{3}{2}}\,\vec{b}^q_{\pi^+\pi^-\eta_8} -
  \frac{1}{\sqrt{2}}\,\vec{b}^q_{\pi^+\pi^-\pi^0} -
  \vec{b}^q_{K^+\pi^-\Kb}\;.
\end{eqnarray}
Only the second relation has all modes measured. (In the $\su$ limit
$\eta_8 = \eta$. However, if the corresponding modes involving $\eta'$
are also measured, the $\eta-\eta'$ mixing can be taken into account
explicitly. See~\cite{Engelhard:2005hu} for more details.)  Using this
relation we can place a bound (experimental values for the branching
ratios \cite{HFAG,Aubert:2001xb} are presented in appendix
\ref{sec:exp-data}):
\begin{equation}
  \label{eq:19}
  \left|\widehat{\xib}_{K^+K^-\K }\right| \leq
  0.22\left(\sqrt{\frac{\frac{3}{2}\mathcal{B}(\pi^+\pi^-\eta_8)}{\mathcal{B}(K^+K^-\K
        )}} +
    \sqrt{\frac{\frac{1}{2}\mathcal{B}(\pi^+\pi^-\pi^0)}{\mathcal{B}(K^+K^-\K
        )}} +
    \sqrt{\frac{\mathcal{B}(K^+\pi^-\Kb)}{\mathcal{B}(K^+K^-\K
        )}}\right) \leq 0.41. 
\end{equation}
This implies that
\begin{equation}
  \label{eq:20}
  \left|\xib_{K^+K^-\K }\right| \leq 0.78.
\end{equation}
Clearly better measurements are needed in order to make this bound
more constraining.

\subsubsection{Dynamical Assumptions}
\label{sec:dynam-assumpt}
One can use simplifying dynamical assumptions by neglecting the effect
of small contributions from exchange, annihilation, and penguin
annihilation diagrams~\cite{Dighe:1995bm}. Such a simplification leads
to new relations. We present relations involving two $\Delta s = 0$
modes:
\begin{eqnarray}
  \label{eq:21}
  &\vec{a}^q_{K^+K^-\K } = -\sqrt{2}\,
  \vec{b}^q_{K^+K^-\pi^0} + \vec{b}^q_{\pi^+K^-\K }\;,\\
  &\vec{a}^q_{K^+K^-\K } = \sqrt{6}\,
  \vec{b}^q_{K^+K^-\eta_8} + \vec{b}^q_{\pi^+K^-\K }\;.
\end{eqnarray}
Upper bounds for the branching ratios of the modes in
\eqref{eq:21} have been obtained. Using these we get
\begin{equation}
  \label{eq:22}
  \left|\widehat{\xib}_{K^+K^-\K }\right| \leq
  0.22\left(\sqrt{\frac{2\mathcal{B}(K^+K^-\pi^0)}{\mathcal{B}(K^+K^-\K )}} + \sqrt{\frac{\mathcal{B}(\pi^+K^-\K )}{\mathcal{B}(K^+K^-\K )}}\right) \leq 0.48.
\end{equation}
This implies that
\begin{equation}
  \label{eq:23}
  \left|\xib_{K^+K^-\K }\right| \leq 1.02,
\end{equation}
a weaker constraint compared to eq.~\eqref{eq:20}. Again, better
measurements may improve this constraint.

\subsection{$B\to K_SK_SK_L$ ($K_LK_LK_S$)}
\label{sec:su-relat-k_sk_sk_l}

The $K_SK_SK_L$ mode, although having the same $\Delta s = 1$
contribution as $K_SK_SK_S$ (namely, $\K \K \Kb$) does not
necessarily have the same $\su$ relations. This is due to the fact
that the $\su$ relations given in \cite{Engelhard:2005hu} depend only
on the symmetric part of the amplitude. The full amplitude
fulfills only a subset of the $\su$ relations given there.

Since the mode $\K \K \Kb$ is symmeric under the exchange of
two of its constituents, the strongest $\su$ bounds come from
modes having that symmetry as well. We define
\begin{eqnarray}
  \label{eq:24}
  &\ket{\mathcal{S}(M_1M_2)M_3}& = \frac{1}{\sqrt {2}}(\ket{M_1M_2M_3} +
  \ket{M_2M_1M_3})\;,\\
  &\ket{\mathcal{S}(M_1M_1)M_2}& = \ket{M_1M_1M_2}\;,\\
  &\ket{\mathcal{A}(M_1M_2)M_3}& = \frac{1}{\sqrt {2}}(\ket{M_1M_2M_3} -
  \ket{M_2M_1M_3})\;,
\end{eqnarray}
where $M_1$, $M_2$ and $M_3$ are taken to be different meson
here. Using this notation, the $\su$ relations take the form
\begin{equation}
  \label{eq:25}
  \vec a_{\K \K \Kb}^q = \sum_{f'=\mathcal{S}(M_1M_2)M_3} X_{f'} \vec b^q_{f'}.
\end{equation}
Noting that
\begin{equation}
  \label{eq:26}
   \braket{K_SK_SK_L}{\frac{1}{\sqrt{3}}(\Kb\K\K + \K\Kb\K - \K\K\Kb)} =
\sqrt{\frac{3}{8}},
\end{equation}
while the two other orthogonal combinations of $\K\K\Kb$,
$\K\Kb\K$ and $\Kb\K\K$ do not contribute, we write
\begin{equation}
  \label{eq:27}
  \vec a^q_{SSL} = \frac{1}{\sqrt{8}}(\vec
  a^q_{K_0\bar K_0K_0} + \vec a^q_{\bar K_0K_0K_0} -\vec a^q_{\K \K \Kb}),
\end{equation}
implying
\begin{equation}
  \label{eq:28}
  \norm{\vec a^q_{SSL}} \leq \frac{3}{\sqrt 8}\norm{\vec a^q_{\K \K \Kb}}.
\end{equation}
Using eqs. \eqref{eq:14}, \eqref{eq:15} and \eqref{eq:16} with
\eqref{eq:25} and \eqref{eq:28} we obtain
\begin{equation}
  \label{eq:29}
  \left|\xibh_{K_SK_SK_L}\right|^2 \leq \frac{9}{8} \abs{\frac{V_{us}}{V_{ud}}}^2
  \frac{\left(\sum\limits_{f'=\mathcal{S}(M_1M_2)M_3}\abs{X_{f'}}\sqrt{ \mathcal{B}(B\to M_1M_2M_3)}
    \right)^2}{\mathcal{B}(B \to K_SK_SK_L)},
\end{equation}
where we used the relation $\norm{\vec
b_{\mathcal{S}(M_1M_2)M_3}^q} \leq \norm{\vec b^q_{M_1M_2M_3}}$.

Since there is no measurement of the branching ratio
$\mathcal{B}(B\to K_SK_SK_L)$, we cannot put numerical bounds at
present. Still, we present the most promising $\su$ relations below.

\subsubsection{Two $\Delta s = 0$ Amplitudes}
\label{sec:four-delta-s} We find two amplitude relations involving
two $\Delta s=0$ modes:
\begin{eqnarray}
  \label{eq:30}
  &\vec{a}^q_{\K  \K \Kb} =
  \sqrt{3}\,\vec{b}^q_{\mathcal{S}(\eta_8\K ) \Kb} -
  \vec{b}^q_{\mathcal{S}(\pi^0\K ) \Kb},\\
  \label{eq:31}
  &\vec{a}^q_{\K  \K  \Kb} =
  \sqrt{3}\,\vec{b}^q_{\mathcal{S}(\eta_8\Kb) \K } -
  \vec{b}^q_{\mathcal{S}(\pi^0 \Kb) \K }\;.
\end{eqnarray}
The modes $X\K \Kb$ with $X\in \{\eta,\pi^0\}$ have not been
measured yet (this requires the measurement of both $XK_SK_S$ ($XK_LK_L$)
and $XK_SK_L$ modes).


\subsubsection{Dynamical assumption}
\label{sec:dynamical-assumption}
Using the dynamical assumption of section~\ref{sec:dynam-assumpt}, we
find the relation
\begin{eqnarray}
  \label{eq:33}
  &\vec{a}^q_{\K  \K  \Kb} = \sqrt{2}\,\vec{b}^q_{\mathcal{S}(\pi^+ \Kb) \K }\;.
\end{eqnarray}

The branching ratio $\mathcal{B}(B\to \pi^+\Kb \K )$ is yet to be measured.

\subsection{$B\to K_SK_SK^+$ ($K_LK_LK^+$)}
\label{sec:bto-k+k_sk_s-k+k_lk}

The state $K_SK_SK^+$ can only result from the state
\begin{equation}
  \label{eq:34}
  \braket{K_SK_SK^+}{\mathcal{S}(\K \Kb) K^+}=\sqrt{\frac{1}{2}}\;,
\end{equation}
The orthogonal combination $\mathcal{A}(\K \Kb)K^+$ does not
contribute.  The table of reduced matrix elements is therefore the
same table used for $K_SK_SK_L$ above.

A nice feature of charged modes is that the $B^+$ is a singlet of the
U-spin subgroup of $\su$. As a thumb rule, U-spin has a good chance of
giving simple relations suitable for our needs since it can change the
strangeness of the final state without changing the charge.

Indeed, for $\K\Kb K^+$ we find the simple U-spin relation
\begin{equation}
  \label{eq:35}
  \vec{a}^q_{\K\Kb K^+} = \vec{b}^q_{\Kb\K \pi^+}\;.
\end{equation}
Since we are interested in the symmetric combination, the
relation~(\ref{eq:35}) translates to
\begin{equation}
  \label{eq:36}
  \vec{a}^q_{K_SK_S K^+} = \vec{b}^q_{K_SK_S \pi^+}\;.
\end{equation}
The bound on $\axibh$ is therefore
\begin{equation}
  \label{eq:37}
  \left|\xibh_{K_SK_SK^+}\right|=\left|\frac{V_{us}}{V_{ud}}\right|
  \sqrt{\frac{\Gamma(B^+\to K_SK_S\pi^+)}{\Gamma(B^+\to K_SK_SK^+)}}
  \leq 0.12\;,
\end{equation}
which leads to
\begin{equation}
  \label{eq:38}
  \left|\xib_{K_SK_SK^+}\right| \leq 0.19\;.
\end{equation}

We note that this bound can further improve if the constraint on
$K_SK_S\pi^+$ will be strengthend.

\subsection{$B^+\to K^+K^+K^-$}
\label{sec:b-k+k+k-}
We have the following relations involving only a single $\Delta s = 0$
amplitude:
\begin{eqnarray}
  \label{eq:39}
  &\vec{a}^q_{K^+K^+K^-} =& \sqrt{2}\vec{b}^q_{\mathcal{S}(\pi^+K^+)K^-},\\
  \label{eq:40}
  &\vec{a}^q_{K^+K^+K^-} =& \vec{b}^q_{\pi^+\pi^+\pi^-}.
\end{eqnarray}

Using \eqref{eq:40} the bound on $\axibh$ is therefore
\begin{equation}
  \label{eq:41}
  \left|\xibh_{K^+K^+K^-}\right|=\left|\frac{V_{us}}{V_{ud}}\right|
  \sqrt{\frac{\Gamma(B^+\to \pi^+\pi^+\pi^-)}{\Gamma(B^+\to K^+K^+K^-)}}
  \leq 0.09\;,
\end{equation}
which leads to
\begin{equation}
  \label{eq:42}
  \left|\xib_{K^+K^+K^-}\right| \leq 0.15\;.
\end{equation}

Note that we use the value of $\mathcal{B}(B^+\to
\pi^+\pi^+\pi^-)$ with resonances removed, and that this bound can
further improve if the constraint on $\mathcal{B}(B^+\to
\pi^+\pi^+\pi^-)$ will be strengthend.

\subsection{$B^+\to K_SK_LK^+$}
\label{sec:b+to-k_sk_lk+}

The state $K_SK_LK^+$ can only result from the state
\begin{equation}
  \label{eq:43}
  \braket{K_SK_LK^+}{\mathcal{A}(\K \Kb) K^+}=-\sqrt{\frac{1}{2}}\;.
\end{equation}
The orthogonal combination $\mathcal{S}(\K \Kb)K^+$ does not
contribute.

There is no measurement yet of the branching ratio
$\mathcal{B}(B^+\to K_SK_LK^+)$. Still, we present the most
promising $\su$ relations below.

\subsubsection{A single $\Delta s = 0$ Amplitude}
\label{sec:one-delta-s}

Since we are interested in the anti-symmetric combination, the
relation~(\ref{eq:35}) translates to
\begin{equation}
  \label{eq:44}
  \vec{a}^q_{K_SK_L K^+} = -\vec{b}^q_{K_SK_L \pi^+}\;.
\end{equation}
The branching ratio $\mathcal{B}(B^+\to K_SK_L \pi^+)$ has not been
measured yet.

\subsubsection{Two $\Delta s = 0$ Amplitudes}
\label{sec:two-delta-s}

We also present relations involving two $\Delta s = 0$ modes:
\begin{eqnarray}
  \label{eq:45}
  &\vec{a}^q_{\mathcal{A}(\K \Kb)K^+} =&
  \sqrt{\frac{3}{8}}\,\vec{b}^q_{\mathcal{A}(\eta_8 \Kb)
    K^+} + \sqrt{\frac{1}{8}}\,\vec{b}^q_{\mathcal{A}(\Kb \pi^0)
    K^+},\\
  \label{eq:46}
  &\vec{a}^q_{\mathcal{A}(\K \Kb)K^+} =&
  \sqrt{2}\,\vec{b}^q_{\mathcal{A}(\Kb \pi^0)
    K^+} + \sqrt{3}\,\vec{b}^q_{\mathcal{A}(\pi^0 \eta_8)
    \pi^+},\\
  \label{eq:47}
  &\vec{a}^q_{\mathcal{A}(\K \Kb)K^+} =&
  \sqrt{\frac{2}{3}}\,\vec{b}^q_{\mathcal{A}(\eta_8 \Kb)
    K^+} - \frac{1}{\sqrt{3}}\,\vec{b}^q_{\mathcal{A}(\pi^0 \eta_8)
    \pi^+}.
\end{eqnarray}
Note that~(\ref{eq:47}) is a linear combination of~(\ref{eq:46})
and~(\ref{eq:45}).

The branching ratios $\mathcal{B}(B^+\to \eta_8 \Kb K^+)$ and
$\mathcal{B}(B^+\to \eta_8 \pi^0 \pi^+)$ have not been
measured yet.


\section{Conclusions}
\label{sec:conclusions} In this work we considered the use of the
approximate $\su$ flavour symmetry of the SM to bound the ratio
between CKM suppressed and CKM favoured terms in $B\to KKK$ decay
amplitudes. This ratio plays an important role in constraining the CP
asymmetries of these modes.

We presented several $\su$ relations that can be used to put bounds on
the asymmetries. For some $B\to KKK$ modes, the current experimental
data is insufficient in order to significantly bound the asymmetries.
For those modes, our work can only provide the most promising
relations that will allow, once experimental data becomes available,
to place stronger bounds on the asymmetries.

For other modes we get the following current constraints:
\begin{eqnarray}
  \label{eq:48}
  &\abs{\xib_{K^+K^-\K }} \leq& 0.78,\\
  &\abs{\xib_{K_SK_SK^+}} \leq& 0.19,\\
  &\abs{\xib_{K^+K^+K^-}} \leq& 0.15.
\end{eqnarray}
Future experimental data can lead to stronger bounds. These can be
confronted with current and future measured CP asymmetries. Currently,
we find all asymmetries to be well within the $\su$ bound.

The work can be extended to include other three body $\Delta s=1$
modes as well. For example, the measured $B\to K^+\pi^+\pi^-$ or $B\to
K^+\pi^-\pi^0$, or states with vector mesons can be considered.

The hope is that, given more and better experimental data, these
decay modes and $\su$ relations will provide us with additional
unambiguous tests of the SM mechanism of CP violation.

\acknowledgments We thank Yossi Nir for helpful discussions and his
comments on the manuscript.

\appendix
\section{Experimental data}
\label{sec:exp-data} We quote experimental data relevant to three
pseudoscalar final states. Measurements where resonant
contributions are removed from the sample are denoted by (NR). The
currently measured $\Delta s=\pm1$ modes are~\cite{HFAG}:
\begin{equation}
  \label{eq:49}
  \begin{aligned}
    \mathcal{B}(K_SK_SK_S) &= (6.2 \pm 0.9) \times
    10^{-6},\\
    \mathcal{B}(K^+ \pi^+ \pi^-) &= (54.1 \pm 3.1) \times
    10^{-6},\\
    \mathcal{B}(K^+ \pi^+ \pi^-)^{\text{(NR)}} &= (2.9^{+1.1}_{-0.9}) \times
    10^{-6},\\
    \mathcal{B}(K^+ K^- K^+) &= (30.1 \pm 1.9) \times
    10^{-6},\\
    \mathcal{B}(K^+ K_S K_S) &= (11.5 \pm 1.3) \times
    10^{-6},\\
    \mathcal{B}(\eta K^+ \pi^-) &= (31.7^{+2.9}_{-3.2}) \times
    10^{-6},\\
    \mathcal{B}(\K  \pi^+ \pi^-) &= (43.8 \pm 2.9) \times
    10^{-6},\\
    \mathcal{B}(K^+ \pi^- \pi^0) &= (35.6^{+3.4}_{-3.3}) \times
    10^{-6},\\
    \mathcal{B}(K^+ \pi^- \pi^0)^{\text{(NR)}} &< 4.6 \times
    10^{-6},\\
    \mathcal{B}(K^+ K^- \K ) &= (24.7 \pm 2.3) \times
    10^{-6},\\
    \mathcal{B}(\K \pi^+\pi^0) &< 66 \times 10^{-6}.
  \end{aligned}
\end{equation}
The currently measured or constrained $\Delta s=0$ modes
are~\cite{HFAG,Eidelman:2004wy,Aubert:2001xb}:
\begin{equation}
  \label{eq:50}
  \begin{aligned}
    \mathcal{B}(\pi^+\pi^-\pi^+) & = 16.2 \pm 1.5,\\
    \mathcal{B}(\pi^+ \pi^- \pi^+)^{\text{(NR)}} &< 4.6 \times
    10^{-6},\\
    \mathcal{B}(\pi^+ \pi^- \eta) &= (6.2^{+2.0}_{-1.7}) \times
    10^{-6},\\
    \mathcal{B}(K^+ K^- \pi^+) &< 6.3 \times
    10^{-6},\\
    \mathcal{B}(K_S K_S \pi^+) &< 3.2 \times
    10^{-6},\\
    \mathcal{B}(K^+ \overline{K}{}^0 \pi^0) &< 24 \times
    10^{-6},\\
    \mathcal{B}(\K K^- \pi^+) &< 21.0 \times
    10^{-6}, \\
    \mathcal{B}(K^+ K^- \pi^0) &< 19 \times
    10^{-6},\\
    \mathcal{B}(K^+ \overline{K}{}^0 \pi^-) &< 18 \times
    10^{-6},\\
    \mathcal{B}(\pi^+ \pi^- \pi^0)^{\text{(NR)}} &< 7.3 \times
    10^{-6}.
\end{aligned}
\end{equation}

\end{document}